# Revisiting Seismicity Criticality: A New Framework for Bias Correction of Statistical Seismology Model Calibrations

**Jiawei Li[1]†, Didier Sornette[1]†, Zhongliang Wu[1,2], Jiancang Zhuang[1,3], and Changsheng Jiang[4]**

[1]Institute of Risk Analysis, Prediction and Management (Risks-X), Academy for Advanced Interdisciplinary Studies, Southern University of Science and Technology (SUSTech), Shenzhen, China.

[2]Institute of Earthquake Forecasting, China Earthquake Administration, Beijing, China.

[3]The Institute of Statistical Mathematics, Research Organization of Information and Systems, Tokyo, Japan.

[4]Institute of Geophysics, China Earthquake Administration, Beijing, China.

Corresponding author: Didier Sornette (didier@sustech.edu.cn); Jiawei Li (lijw@cea-igp.ac.cn);

†These two authors contributed equally.

**Key Points:**

- Identification and quantification of boundary effects, finite-size effects, and censorship that bias the calibrations of statistical seismology models
- Development of a framework for correcting biases in the branching ratio $n$ of the ETAS model from the estimated dependence of the apparent branching ratio $n_{\mathrm{app}}(M_{\mathrm{co}})$ as a function of changing cut-off magnitudes $M_{\mathrm{co}}$
- After corrections to obtain $n_{\mathrm{true}}$, seismicity is still found subcritical ($n_{\mathrm{true}} < n_c = 1$), and many small triggered earthquakes may not be fertile.

## Abstract

The Epidemic-Type Aftershock Sequences (ETAS) model and its variants effectively capture the space-time clustering of seismicity, setting the standard for earthquake forecasting. Accurate unbiased ETAS calibration is thus crucial. But we identify three sources of bias, (i) boundary effects, (ii) finite-size effects, and (iii) censorship, which are often overlooked or misinterpreted, causing errors in seismic analysis and predictions. By employing an ETAS model variant with variable spatial background rates, we propose a method to correct for these biases, focusing on the branching ratio $n$, a key indicator of earthquake triggering potential. Our approach quantifies the variation in the apparent branching ratio ($n_{\mathrm{app}}$) with increased cut-off magnitude ($M_{\mathrm{co}}$) above the optimal cut-off ($M_{\mathrm{co}}^{\mathrm{best}}$). The $n_{\mathrm{app}}(M_{\mathrm{co}})$ function yields insights superior to traditional point estimates. We validate our method using synthetic earthquake catalogs, accurately recovering the true branching ratio ($n_{\mathrm{true}}$) after correcting biases with $n_{\mathrm{app}}(M_{\mathrm{co}})$. Additionally, our method introduces a refined estimation of the minimum triggering magnitude ($m_0$), a crucial parameter in the ETAS model. Applying our framework to the earthquake catalogs of California, New Zealand, and the China Seismic Experimental Site (CSES) in Sichuan and Yunnan provinces, we find that seismicity hovers away from the critical point, $n_c = 1$, remaining distinctly subcritical, however with values tending to be larger than recent reports that do not consider the above biases. It is interesting that, $m_0$ is found around 4 for California, 3 for New Zealand and 2 for CSES, suggesting that many small triggered earthquakes may not be fertile. Understanding seismicity's critical state significantly enhances our comprehension of seismic patterns, aftershock predictability, and informs earthquake risk mitigation and management strategies.

## Plain Language Summary

The Epidemic-Type Aftershock Sequences (ETAS) model, a useful tool for earthquake forecasting, captures seismicity clustering effectively. However, biases from boundary effects, finite-size effects, and censorship often distort its accuracy. We develop a method employing an ETAS model allowing variable spatial background rates to correct these biases, especially focusing on the branching ratio $n$, indicative of earthquake triggering potential. This method, validated with synthetic catalogs, precisely adjusts the apparent branching ratio $n_{\mathrm{app}}$ based on increased cut-off magnitude $M_{\mathrm{co}}$ to the true branching ratio $n_{\mathrm{true}}$, revealing more insightful estimates. It also refines the estimation of the minimum triggering magnitude $m_0$, crucial for ETAS models. Application to catalogs from California, New Zealand, and China Seismic Experimental Site (CSES) confirms that seismicity is subcritical, with $m_0$ estimates indicating many small earthquakes are not potent triggers. This insight into seismicity's critical state advances our understanding of earthquake patterns, aftershock predictability, and guides risk mitigation strategies.

## 1 Introduction

Seismicity and fault rupturing in the Earth's crust can be likened to epidemic processes, as exemplified by the Epidemic-Type Aftershock Sequences (ETAS) model (Kagan & Knopoff, 1981; 1987; Kagan 1991; Ogata, 1988; 1998; Musmeci & Vere-Jones, 1992; Ogata & Zhuang, 2006), wherein background earthquakes (“immigrants”), assumed to be driven by the forces of plate tectonics, have the potential to trigger cohorts of earthquakes, referred to as first-generation “daughters.” These first-generation events, possessing fertility, can in turn act as “mothers,” triggering second-generation “daughters” and creating a cascading effect (Hawkes & Oakes, 1974; Helmstetter & Sornette, 2002; Felzer et al., 2002; 2003; Zhuang et al., 2002; 2005).

In this dynamic process, a crucial parameter known as the branching ratio $n$ plays a pivotal role in reflecting the state of the crustal system. This parameter signifies both the mean number of fertile first-generation "daughters" triggered per "mother" and the fraction of all triggered events relative to the total number of earthquakes. The transition point occurs at $n = n_c = 1$, demarcating the subcritical regime ($n < 1$), where the sequence of events is stationary, from the supercritical regime ($n > 1$), where the number of earthquakes increases exponentially over time with a finite probability (Helmstetter & Sornette, 2002). Consequently, the branching ratio $n$ serves as an indicator of the proportion of observed earthquakes triggered by preceding events when the magnitude distribution can be separated from the other components (Zhuang et al. 2013). The extent to which $n$ falls below the critical point $n_c$ has significant implications on our understanding of seismicity patterns, on the forecastability of aftershocks and for earthquake hazard mitigation and risk management. More profoundly, since the introduction (Bak et al., 1987) and application of the concept of self-organized criticality to seismicity (Bak & Tang, 1989; Sornette & Sornette, 1989), a flurry of studies have proposed that earthquake processes might operate at or near a critical point (Fisher et al., 1997; Zöller et al., 2001; Keilis-Borok & Soloviev, 2003; Main et al., 2006; Shcherbakov et al., 2006; Wanliss et al., 2017; de Arcangelis et al., 2016), in the sense of critical phenomena in Statistical Physics. In contrast, research by Wu (1998) and subsequent findings in statistical seismology suggest that seismicity might be a self-organized process that does not reach a critical state, with evidence showing deviations from the critical point at $n_c = 1$ (Chu et al., 2011; Nandan et al., 2021b; 2022). On the other hand, Huang et al. (1998) introduced a generalized sandpile model on a hierarchical system of faults, suggesting the possible coexistence between self-organised criticality and genuine criticality of individual large earthquakes. Whether or not they are close to such criticality is frequently seen as having profound implications for our understanding of the physics behind earthquakes. Hence, obtaining an unbiased estimation of the branching ratio $n$ of seismicity holds significant importance.

A few previous authors have acknowledged the existence of several sources of biases in the estimation of $n$, which are non-stationarity and non-uniformity of the background rate when not properly accounted for properly, spatial and temporal boundary effects, finite-size effects and censorship. In the present study, our primary objective is to quantify the biases in the calibration of the branching ratio using the ETAS model. We develop a method to correct the apparent branching ratio $n_{app}$ obtained from calibrations, in order to derive a good approximation of the true branching ratio $n_{true}$, via a procedure validated through synthetic tests. Our methodology leverages the optimal estimation of the cut-off magnitude ($M_{co}^{best}$) for a given catalog. We focus on how $n_{app}$ shifts when the used cut-off magnitude $M_{co}$ is artificially increased above $M_{co}^{best}$, enabling us to methodically investigate the impacts of boundary effects, finite-size effects of seismicity sample and censorship. Applying this framework to earthquake catalogs from California, New Zealand, and the China Seismic Experimental Site (CSES), we re-examine the question of whether seismicity operates near a critical point ($n_c = 1$). This reassessment carries significant implications for our understanding of seismic patterns, improving aftershock predictability, and shaping strategies for earthquake risk mitigation and hazard management.

The organization of the manuscript starts with Section 2, which presents the ETAS model, defines the different estimation methods of its branching ratio $n$ and explains the meaning of subcriticality, criticality and supercriticality. Section 2 also describes the different effects that can bias the estimation of $n$ and reviews existing methods for correcting these biases. Section 3 presents synthetic tests demonstrating the biases and present our proposed correction method, which is illustrated and tested on synthetic catalogues. Section 4 describes the empirical catalogues used in this study. Section 5 examines in detail the application of our correction method to the empirical data of California, New Zealand and the China Seismic Experimental Site (CSES) in Sichuan and Yunnan provinces. Section 6 discusses the results and concludes.

## 2 The ETAS model, calibration and biases

### 2.1 The ETAS model

The ETAS model is an instantiation adapted to seismicity of the self-excited Hawkes stochastic point process, which is non-Markovian when the temporal memory kernel is non-exponential (Hawkes & Oakes, 1974; Kagan & Knopoff, 1981; 1987; Ogata, 1988; 1998). It enjoys two equivalent interpretations:

1) A simple ensemble branching model (Hawkes & Oakes, 1974; Kagan 1991; Sornette & Werner, 2005b). In this conceptualization, a given sequence of events is mapped on many trees, each tree corresponding to a specification of what are the background events and what are the "mothers" and triggered "daughters." Each tree starts with a "immigrants" (or background earthquake) and all the branches that emanate from this "immigrants" correspond to the aftershocks and then aftershocks of aftershocks and so on. Each tree representation is endowed with a probability representing the likelihood of this specific association between earthquakes, given the observed sequence. The tree probabilities derive from the expression of the intensity of the ETAS model given by Formula (1) below (Zhuang et al., 2002). In this representation, the ETAS model comprises statistically independent Poisson seismicity clusters (each cluster being a tree), while there is dependence within every cluster (tree branches corresponding to all generations). Consequently, the branching ratio *n* is an average, encompassing not only magnitudes but also across an ensemble of realizations of the stationary point process (for $n \leq 1$) sustained by the presence of background events.

2) A coupled earthquake interaction model (Helmstetter & Sornette, 2002). In this conceptualization, each non-background event is collectively triggered by all preceding events, with each event contributing a weight determined by the fertility law (or productivity law) $F(m)$ that decays in space and time governed by the space kernel $S(x, y, m)$ and time kernel $T(t)$. The product of $F(m)$, $S(x, y, m)$, and $T(t)$ forms the triggering function. Consequently, at any given spatiotemporal point $(x, y, t)$, the ETAS model is defined by the seismicity rate (or intensity) $\lambda$, which encapsulates the dependence on the historical seismicity $H_t$ recorded until time $t$, as:

$$\lambda(t, x, y|H_t) = \mu(x, y) + \sum_{i:t_i<t} F(m_i)T(t - t_i)S(x - x_i, y - y_i, m_i) \quad (1)$$

The seismicity rate, at time $t$ and position $(x, y)$, is thus the sum of contributions from both the background intensity function, $\mu(x, y)$, and a sum over all preceding earthquakes, in which each past earthquake contributes additivity with a multiplicative separable spatio-temporal kernel and with an amplitude given by its fertility $F(m)$. In this formulation, the branching ratio *n* can be interpreted as the contribution of a past earthquake to a future earthquake, averaged over an ensemble of realizations and over all magnitudes. This perspective is the only one possible for nonlinear generalisations of the Hawkes model whose triggering functions depend nonlinearly on previous events (Ouillon & Sornette, 2005; Sornette & Ouillon, 2005; Kanazawa & Sornette, 2021; 2023).

The rigorous mathematical demonstration of the equivalence of these two representations was first presented in Hawkes & Oakes (1974). Sornette & Werner (2005b) provided an intuitive understanding of the underlying mechanism for this equivalence, based on the formulation of the seismicity rate of the ETAS model as a linear sum over past earthquakes combined with the exponential form of the hazard rate of point process.

Formula (1) for the seismicity rate is complemented by the Gutenberg-Richter distribution given by

$$P(m) = \frac{b\ln(10)10^{-bm}}{10^{-bm_*} - 10^{-bm_{\max}}}, m_* \leq m \leq m_{\max} \tag{2}$$

where $b$ is typically close to one. In this formula, $m_{\max}$ is the largest magnitude of earthquakes capable of triggering other earthquakes and thus the Gutenberg-Richter distribution is truncated at $m_{\max}$. The lower magnitude threshold $m_*$ is the minimum magnitude of an earthquake that can be triggered by prior events. It is important to stress that the ETAS model also requires an additional characteristic magnitude $m_0$, which is the smallest magnitude of earthquakes capable of triggering other earthquakes. In other words, the fertility of an earthquake is zero if its magnitude is smaller than $m_0$. It plays the role of an "ultra-violet" cut-off ensuring the convergence and stationarity of the model (Sornette & Werner, 2005a). The fertility law defining the number of direct aftershocks ("daughters") of a "mother" event of magnitude $m$ is thus given by:

$$F(m) = K\exp[\alpha(m - m_0)], m_0 \leq m \leq m_{\max} \tag{3}$$

where $K$ and $\alpha$ are constants. Most previous publications have assumed explicitly or implicitly that $m_* = m_0$ but nothing prevents in principle that $m_* < m_0$, i.e., many earthquakes are triggered so small that they are not fertile and they do not contribute to future seismicity.

The other terms in the sum in the r.h.s. of Formula (1) read

$$T(t - t_i)S(x - x_i, y - y_i, m_i) = \frac{T_{\text{norm}}e^{-\frac{t-t_i}{\tau}}}{(t - t_i + c)^{1+\omega}} \cdot \frac{S_{\text{norm}}}{\left[(x - x_i)^2 + (y - y_i)^2 + de^{\gamma(m_i - M_{co})}\right]^{1+\rho}} \tag{4}$$

which are respectively the Omori-Utsu law $T(t - t_i)$ defined with the constants $c$, $\omega$, and $\tau$, and the spatial Green function with constants $d$, $\gamma$, and $\rho$, which describe the temporal and spatial dependencies within the model. $T_{\text{norm}}$ and $S_{\text{norm}}$ represent normalization constants for the time and space kernels, ensuring they are proper probability density functions (PDFs).

In this study, the spatial variability of background rates $\mu(x, y)$ is treated as a non-parametric function. Its estimation is through a weighted kernel function used in (Nandan et al., 2021b; 2022),

$$\mu(x, y) = \frac{1}{T}\sum_{i=1}^{N} IP_i \cdot \frac{QD^{2Q}}{\pi} \frac{1}{[(x - x_i)^2 + (y - y_i)^2 + D^2]^{1+Q}} \tag{5}$$

which is a sum over all earthquakes weighted by their probabilities $IP_i$ to be background events (Zhuang et al., 2002). The normalization by the duration $T$ of the primary catalog ensures that $\mu(x, y)$ represents the seismicity rate per unit time, while the factor $\pi^{-1}QD^{2Q}$ ensures normalization per unit area at location $(x, y)$. The choice of the power-law kernel is motivated by prior research indicating its superior performance compared to previously more commonly used Gaussian kernels (e.g., Helmstetter et al., 2007; Nandan et al., 2021a). The estimation of the model, including model parameters $\{Q, D, K, \alpha, c, \omega, \tau, d, \gamma, \rho\}$ and nonparametric spatial variability of the background rate, $\mu(x, y)$, is accomplished through the extended expectation-maximization (EM) algorithm (Veen & Schoenberg, 2008). We refer to Nandan et al. (2021b; 2022) for details.

This ETAS model has been implemented on earthquake catalogs, including those recorded globally, in California and in New Zealand (Nandan et al., 2021b; 2022). Nandan et al. (2021b; 2022) have conducted comparative assessments against other ETAS model variants through pseudo-prospective forecasting experiments to assess its superior performance in earthquake forecasting, which further advances the ETAS model as a benchmark for evaluating alternative earthquake forecasting models (Ogata, 2017; Nandan et al., 2021a; Kamer et al., 2021).

2.2 Definition and estimation methods of the branching ratio $n$

Three methods suggested in prior literature can be used to estimate $n$.

***(1) The formulaic approach.*** The branching ratio is, by definition, the average number of fertile "daughter" per "mother." Given the fertility law presented by Formula (3), the branching ratio is the average of $F(m)$ over all possible values of $m \geq m_0$, weighted by the Gutenberg-Richter (GR) law given by Formula (2). As noted above, preceding publications commonly assume that $m_*$ is equal to $m_0$, i.e., all earthquake are fertile. With $m_* = m_0$, the formula for the branching ratio reads (Sornette & Werner, 2005b)

$$n \equiv \int_{m_0}^{m_{\max}} P(m)F(m)\mathrm{d}m = \begin{cases} \frac{Kb[1-10^{-(b-a)(m_{\max}-m_0)}]}{(b-a)[1-10^{-b(m_{\max}-m_0)}]} & a \neq b \\ \frac{Kb\ln(10)(m_{\max}-m_0)}{1-10^{-b(m_{\max}-m_0)}} & a = b \end{cases} \tag{6}$$

The determination of the branching ratio $n$ thus necessitates knowledge of the parameters $b$, $m_0$, $m_{\max}$, $K$, and $a = \alpha/\ln(10)$. In practical scenarios involving catalogs with a total of $N$ earthquakes with magnitudes $\{m_1, m_2, ..., m_N\}$, Seif et al. (2017) suggested substituting the $P(m)$ in Formula (2) with the empirical frequency-magnitude distribution, leading to the following estimator

$$n = \frac{1}{N}\sum_{i=1}^{N} K \exp\left[\alpha(m_i - m_0)\right] H(m_i - m_0) \tag{7}$$

where H[.] is the Heaviside function. Consequently, this approach simplifies the estimation of the branching ratio $n$ to rely solely on the parameters $N$, $m_0$, $K$, and $a$.

***(2) The counting approach.*** Helmstetter & Sornette (2003), Zhuang et al. (2013) and Seif et al. (2017) demonstrated that the branching ratio $n$ is equivalently represented as the ratio of triggered events to the total number of earthquakes in the ETAS model when $n < 1$. This relationship is expressed by the formula

$$n = \frac{N_{\text{tri}}}{N_{\text{tri}} + N_{\text{bkg}}} \tag{8}$$

where $N_{\text{tri}}$ and $N_{\text{bkg}}$ denote the number of triggered and background events of magnitude larger than or equal to $m_0$, respectively. The calculation of $n$ requires knowledge of either $N_{\text{tri}}$ or $N_{\text{bkg}}$ alone, along with the condition of knowing the total number of earthquakes $N$ of magnitude larger than or equal to $m_0$ ($N = N_{\text{tri}} + N_{\text{bkg}}$). Additionally, Seif et al. (2017) proposed a method for estimating the branching ratio $n$ by considering the proportion of aftershocks that are second or higher generations. This approach is represented as

$$n = \frac{N_{\mathrm{tri}}^{\mathrm{2nd\ +\ higher}}}{N_{\mathrm{tri}}} \tag{9}$$

where $N_{\mathrm{tri}}^{\mathrm{2nd\ +\ higher}}$ is the number of aftershocks of magnitude larger than or equal to $m_0$ that are of generation larger than or equal to 2. In other words, $N_{\mathrm{tri}}^{\mathrm{2nd\ +\ higher}}$ is the number of aftershocks that are themselves triggered by previous aftershocks.

***(3) The mean-variance-based estimation approach.*** Hardiman & Bouchaud (2014) introduced a parameter-independent method for approximating the branching ratio $n$. This technique relies exclusively on the mean ($\bar{N}$) and variance ($\sigma^2$) of the event count within a suitably large time window, as indicated by their sample estimates:

$$\bar{N} = \frac{1}{W}\sum_{i=1}^{W} N(i) \tag{10}$$

$$\sigma^2 = \frac{1}{W-1}\sum_{i=1}^{W}[N(i) - \bar{N}]^2 \tag{11}$$

where $N(i)$ represents the event count in the $i$th time window out of a total of $W$. The branching ratio can be approximated as (Hardiman & Bouchaud, 2014)

$$n \approx 1 - \sqrt{\frac{\bar{N}}{\sigma^2}} \tag{12}$$

This method has been tested using simulated data and empirical financial market data. It is found to be fragile in the presence of a non-constant and/or non-uniform distribution of background events.

2.3 Subcriticality, criticality, or supercriticality of seismicity

In statistical seismology, the branching ratio $n$ is primarily studied using the formulaic and counting approaches. Seif et al. (2017) calibrated the ETAS model with a spatially varying background rate, yielding a branching ratio close to or slightly larger than 1 in Southern California and Italy by using Formula (6). By also allowing for spatial variation but no time dependence of the background seismicity rate, and estimating it within spatial Voronoi partitions, Nandan et al. (2017) estimated the ETAS parameters for California based on earthquakes that occurred from 1981 to 2015. Through the formulaic determination of the branching ratio, the spatial distributions of $K$, $a$, and $b$ lead to significant spatial heterogeneity in branching ratios across California, with values ranging from 0 to 1.2. Nandan et al. (2017) identified regions of California with $n > 1$ while other regions are characterized by $n < 1$, suggested to be associated with respectively negative and positive thermal anomalies. Chu et al. (2011), utilizing an ETAS model with a constant background rate and the formulaic approach, calculated a branching ratio below 1 for global regions with varied tectonic backgrounds. In more recent studies, Nandan et al. (2021b; 2022) developed an ETAS model with an improved parametric representation of the spatial variation of the background seismicity rate. Using the formulaic approach, they computed the branching ratio for global, California, and New Zealand, finding subcritical seismicity, i.e., $n < 1$. In contrast, they show that assuming a uniform background leads to estimated $n$ close to or equal to 1 in the three regions, which would lead to conclude that seismicity is operating at a critical point. In fact, they stress that this last conclusion is erroneous. Because the spatial heterogeneity of the background seismicity is not accounted for adequately when forcing a uniform background on the ETAS calibration, the calibration of the ETAS model is driven to estimated *n* values close to 1, which is the domain of parameters for which the ETAS model produces the strongest spatial and temporal clustering. In other words, criticality is only apparent and is spuriously inferred in the ETAS model calibration as being the statistical bias needed for the mis-specified ETAS model to account for the pre-existing clustering of the underlying background seismicity.

It is useful to summarize the insights obtained from another field (financial time series) in which the effect of non-constancy of the background rate also distorts significantly the estimation of the branching ratio, thus showing that the effect is common to both the spatial and temporal domains. Wheatley et al. (2019) and Wehrli et al. (2021) documented the same effect in the time domain, i.e., when imposing a constant background in the ETAS calibration (in the field of finance as well as in mathematics, the model is called the Hawkes model) of a sequence of events with non-constant background. This question arises in the determination of whether financial markets operate or not in a critical state. The controversy started with the report of Filimonov & Sornette (2012) that the branching ratio (now only using sequences of events in the time domain) of financial time series is clearly less than 1, disqualifying criticality. Shortly after, Hardiman et al. (2013) countered with their finding that *n* is very close to 1, confirming criticality. Among others, the main difference between the two works is the length of the calibrated time series: Filimonov & Sornette (2012) used time series of 10 to 30 minutes durations (with typical waiting times between events of the order of seconds) in order to minimize biases from non-stationarity. In contrast, Hardiman et al. (2013) used much longer time series of up to several months, with the rationale of being better able to reveal a potential critical behavior expressing itself at large temporal scales. Both works calibrated the temporal ETAS model with a constant background. This is where the works of Wheatley et al. (2019) and Wehrli et al. (2021) illuminated the debate. Because financial time series are highly non-constant in time, with high volatility at the opening and at the closing of each day of trading, and with low volatility at lunch time, calibrating such time series extending over several days with an ETAS model constrained to have a constant background pushes spuriously the branching ratio to the critical value 1, so as to model the unaccounted for strong non-stationarity. By using a sophisticated parameterization of the non-constant background with the expectation-maximization (EM) method applied to long time series of several months as in Hardiman et al. (2013), Wheatley et al. (2019) and Wehrli et al. (2021) ruled out in favor of the conclusion of Filimonov & Sornette (2012). It is now understood that, by using short time series, the estimation of the later authors was not susceptible to the bias induced by non-stationarity.

### 2.4 Effects biasing the branching ratio *n*

It is important to note that the three existing methods for estimating the branching ratio have different pros and cons. In comparison with the other two methods, the mean-variance-based estimation approach significantly simplifies the estimation of the branching ratio. However, Wheatley et al. (2019) and Wehrli et al. (2021) demonstrated that this method gives strongly misleading estimations of the branching ratio when the true background rate is non-constant, similarly to the bias identified by Nandan et al. (2021b) when the background rate is spatially non-uniform, as discussed above. Additionally, the counting approach and the mean-variance-based estimation approach are only suitable for $n < 1$. While the formulaic approach is theoretically applicable in any case, its reliability is highly contingent on accurately estimating parameters $b$, $K$, and $a$, along with reasonable assumptions for $m_0$ and $m_{max}$. In contrast, the counting approach relies on accurate estimates of $N_{tri}$ or $N_{bkg}$. The differences in the branching ratio estimations based on these methods are elaborated in Section 5.1.

In addition to the biases induced by non-stationarity and non-uniformity of the background rate, the three aforementioned estimation methods generally neglect or incorrectly address three crucial factors.

1) The boundary effect arises from the use of earthquake catalogs with limited duration and spatial extent. This can lead to a shift in estimation of $n$, as discussed in works by Wang et al. (2010) and Seif et al. (2017), and is exemplified in another context in the line-percolating system depicted in Figure 4 of Vanneste et al. (1991);

2) The finite-size effect stems from the inherently limited statistical size of the seismicity sample. This limitation can result in both large variance and bias in the estimation of $n$, as shown in studies by Sornette & Utkin (2009) and Seif et al. (2017), and is also illustrated in the line-percolating system shown in Figure 3 of Vanneste et al. (1991); and

3) Censorship refers to the missing events that are not taken into account to calibrate the ETAS model due to being too small to be measured but which do trigger earthquakes. Recall that, in the ETAS model, all earthquakes of magnitude larger than a minimum magnitude $m_0$ can trigger "daughters" and thus impact the whole seismicity. This $m_0$ has no reason to be equal to the catalog cut-off magnitude $M_{co}$ above which the catalog is deemed to be complete (Sornette & Werner, 2005b; Saichev & Sornette, 2006;Seif et al., 2017) and is analyzed. Indeed, in the ETAS model, $m_0$ is supposed to be a physical property of earthquake interactions. In contrast, $M_{co}$ strongly depends on and is usually larger than the magnitude completeness $M_c$, which is constrained by instruments, i.e., by the number and coverage of seismic stations and their sensitivity (e.g., Li et al., 2023). As larger investments in more numerous stations are made and better technology is developed, $M_c$ is continuously pushed to smaller values, see e.g. Feng et al. (2022) and Li et al. (2023). Then, it is likely that $m_0$ is smaller than $M_{co}$, which leads to the censorship bias.

Recall that the assumption of a finite value for $m_0$ is crucial for ensuring that seismicity remains bounded in the ETAS model and in most of its variants (Sornette & Werner, 2005a). When $m_0$ is finite and when the Gutenberg-Richter and fertility laws are expressed consistently together with $m_0$ being their lowest magnitude of validity, then the branching ratio $n$ is independent of $m_0$. On the other hand, if the lowest magnitude at which both Gutenberg-Richter and fertility laws apply goes to minus infinity, the branching ratio diverges (is no more defined). Indeed, if $m_0$ (and $m_*$) is pushed towards more and more negative values (smaller and smaller corresponding energies), the number of very small earthquakes grows without bound and the ETAS is no more well-defined in the standard regime where the base-10 fertility exponent $a = \alpha/\ln(10)$ is smaller than the Gutenberg-Richter $b$-value.

The boundary effect and censorship both contribute to the omission of earthquakes that should be considered in a correct calibration of ETAS model. Due to the boundary effect, earthquakes outside the spatio-temporal domain of investigation (the so-called primary catalog) are not considered, while they can be potential “mothers” and “daughters” of earthquakes in the primary catalog. Censorship refers to the fact that all earthquakes of magnitudes between $m_0$ and $M_{co}$ are also missing in the calibration of the ETAS model. But they are also potential "mothers" and "daughters" of the detected earthquakes with $m \geq M_{co}$. These effects disrupt the determination of possible triggering relationships between earthquakes, introducing biases and increasing variance (e.g., finite-size effect) in parameter estimation, particularly in branching ratio estimation (Sornette & Werner, 2005b; Saichev & Sornette, 2006; Sornette & Utkin, 2009; Seif et al., 2017).

In the presence of these effects, the branching ratios that have been estimated in the literature can be considered to be all biased, giving apparent branching ratios $n_{app}$ that are likely different, and perhaps very different, from the true unknown branching ratio $n_{true}$ (Figure 1). This may lead to potentially strongly misleading inferences on the nature of crustal seismicity. The boundary effects is a primary reason for obtaining estimated apparent branching ratios $n_{app}$ less than 1, when using the counting and mean-variance-based estimation methods. In the comparison between $n_{true}$ and $n_{app}$ estimated using the mean-variance-based estimation approach, Hardiman & Bouchaud (2014) demonstrated in their Figure 1, using simulated data, that as $n_{true}$ approaches 1, $n_{app}$ starts to deviate significantly from $n_{true}$. This result is primarily influenced by the boundary effect in the time domain, as their simulation had no spatial component and they were modeling data over a time length of 100,000 seconds. The impact of censorship is also substantial. Chu et al. (2011) and Nandan et al. (2021b) obtained a branching ratio far below 1 for the global catalog, primarily because they both selected $M_{co} = 5$, leading to results likely to be significantly influenced by censorship (Sornette & Werner, 2005b; Saichev & Sornette, 2006; Seif et al., 2017), as we are going to demonstrate quantitatively below.

2.5 Existing methods to correct the apparent branching ratio $n_{app}$

A few previous authors have acknowledged the impact of the boundary effect, the finite-size effect and censorship and have proposed correction procedures for branching ratio estimation. Sornette & Werner (2005b) hypothesized that earthquakes affected by censorship, due to the absence of their "mother" earthquakes, would be classified as background events. Assuming a known $m_0$ and a complete catalog above $M_{co}$, they introduced a censorship correction relating $n_{app}$ from the counting approach to $n_{true}$, formulated as follows:

$$n_{\text{true}} = \begin{cases} n_{\text{app}}\left[\frac{10^{(b-a)(m_{\max}-m_0)}-1}{10^{(b-a)(m_{\max}-M_{\text{co}})}-1}\right] & a \neq b \\ n_{\text{app}}\left[\frac{m_{\max}-m_0}{m_{\max}-M_{\text{co}}}\right] & a = b \end{cases} \tag{13}$$

This formula is such that $n_{app} = n_{true}$ for $m_0 = M_{co}$. Seif et al. (2017) also recognized biases in branching ratio estimation due to censorship and boundary effects. They proposed a formula to derive $n_{true}$ from $n_{app}$ obtained from the formulaic and counting approaches in the presence of the boundary effect in the time domain, which reads:

$$n_{\text{true}} = \frac{n_{\text{app}}}{1 - \delta_{\text{TBE}}} \tag{14}$$

Here $\delta_{TBE}$, derived analytically, represents a boundary effect correction factor in the time domain. This factor is dependent on the Omori parameters of the ETAS model and the temporal duration of the catalog. However, their analysis did not explore boundary effect in the spatial domain and failed to differentiate the distinct effects of boundary and finite-size of seismicity sample on $n_{app}$. Notably, finite-size effects primarily impact the standard deviation of estimations. While Seif et al. (2017) recognized the impact of censorship on branching ratio estimation, they did not propose any correction method for this issue. Moreover, their simulation experiments, exploring $n_{app}$'s variation with $M_{co}$ in simulated catalogs, assumed branching ratios as high as $n = 6$ for $M_{co} = 2.5$ and $n = 25$ for $M_{co} = 2$, contradicting realistic physical constraints that prevent excessively high values to avoid explosive outcomes. Recall that a stationary ETAS process requires $n \leq 1$.

While the corrections proposed by Sornette & Werner (2005b) and Seif et al. (2017) seem sensible, they fail to recognize the impact of a fundamental property of the ETAS model on its calibration: clustering in space and time. Specifically, Werner (2008) pointed out in his Chapter 4 that calibrating the ETAS for seismicity with $M_{co} > m_0$ gives smaller changes of $K$ (and thus of $n$) than predicted by Sornette & Werner (2005b). Indeed, the calibration of ETAS models is in fact controlled by the existence of spatio-temporal clustering, rather independently of the specific genealogy of which earthquake triggers what earthquake. In other words, triggered earthquakes that lose their "mothers" due to their unobservability (via boundary effects and/or censorship) may not be systematically categorized as background but may be correctly seen as part of a triggered cluster. This is due to the presence of other earthquakes in the same family tree within the cluster. To stress this fundamental property again, the genealogy within an earthquake sequence generated by the ETAS model is intrinsically stochastic since the mapping of the ETAS model to branching processes holds in the sense of an ensemble of trees where each possible filiation occurs with probabilities determined from the triggering kernels of the ETAS model. This makes the specific attribution of "motherhood" mostly irrelevant for the problem of parameter estimation. The key concept is that ETAS models account for clustering and only clustering.

## 3. Synthetic tests of the bias in branching ratio *n*

We now present simulations of synthetic catalogs to identify and quantify the three types of biases affecting the branching ratio $n$: boundary effects in time and space, finite-size effects, and censorship. Employing the ETAS model defined in Section 2.1, we generate synthetic catalogs for the study regions (Table S1). In the present study, the spatial and temporal extent used for synthesizing the earthquake catalogs is identical to that of the primary catalog (Figures S1-S3 and Table S1). We simulate synthetic catalogs for the study regions, selecting parameters based on insights gained from subsequent calibrations performed on real catalogs. To simulate a variety of scenarios, we vary the parameters $K$ and $\alpha$ of the fertility law within the ranges of 0.2085–0.9931 and 0.0084–0.4601, respectively, allowing us to model earthquake catalogs under different $n_{true}$. Utilizing the calibrated $Q$, $D$, and $IP$ at $M_{co}^{best}$ and using Formula (4) for the background rate, we generate the probability density function $PDF_{\mu(x,y)}$ for background earthquakes. The number of background earthquakes is determined using the Gutenberg-Richter relationship fitted above $M_{co}^{best}$ to calculate the total expected number of earthquakes with magnitude greater than or equal to $m_0$, and then using the intended $n_{true}$ to calculate the expected number of background earthquakes within that total. With this number as the parameter for a Poisson distribution, 100 random background earthquake counts are generated, resulting in 100 sets of background earthquake catalogs for each $n_{true}$. These synthesized background earthquakes, along with the observed earthquake catalogs, trigger cascading earthquakes according to the laws of the ETAS model.

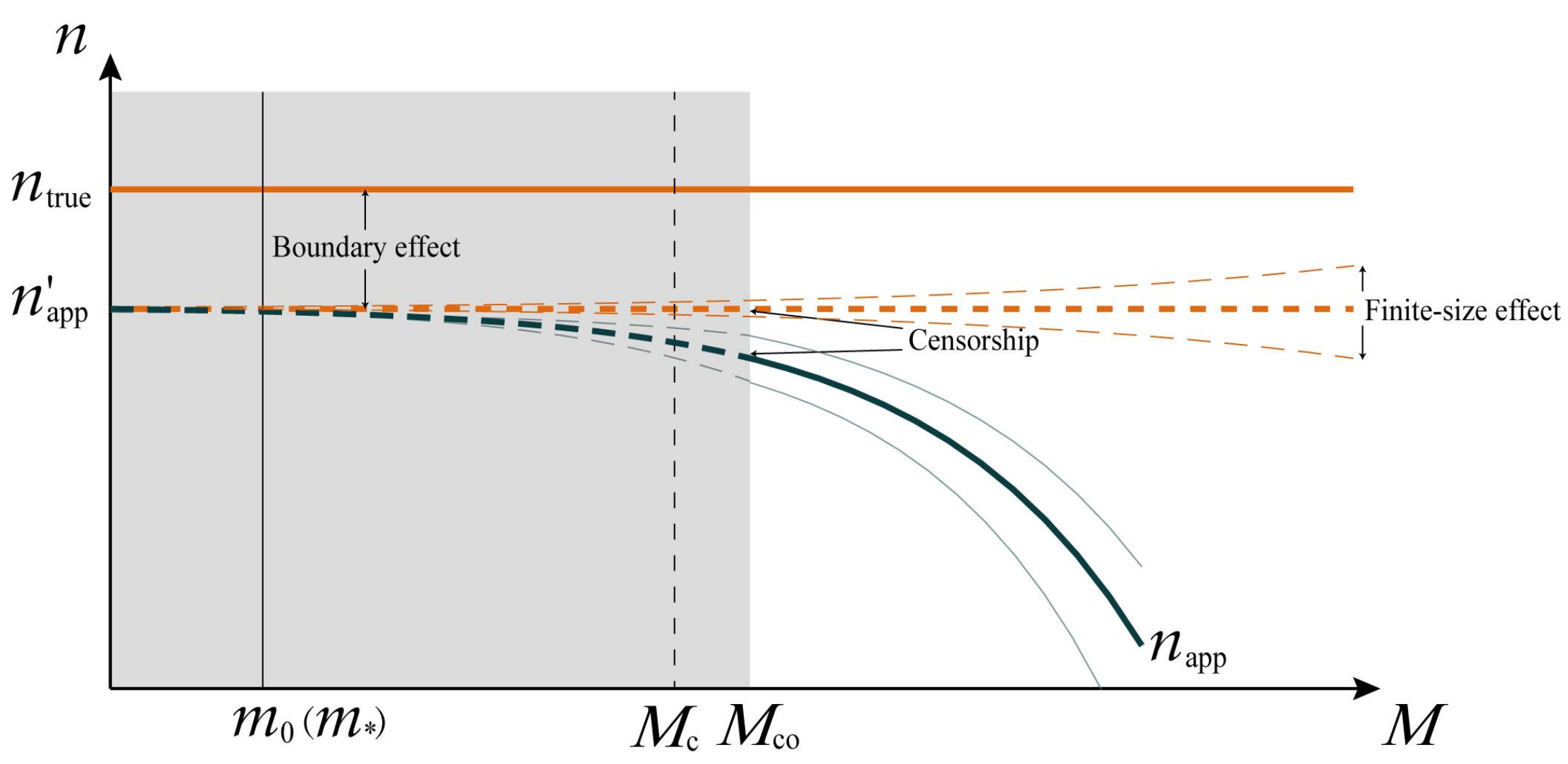

**Figure 1**. Schematic diagram illustrating the biases that transform the true branching ratio ($n_{\text{true}}$) into the apparent branching ratio ($n_{\text{app}}$). This diagram abstracts from common features observed in simulation experiments. In these experiments, $n_{\text{true}}$ (orange solid line) is downwardly biased to $n'_{\text{app}}$ (orange dashed line) due to the boundary effect in time and/or space, with its standard deviation (orange dashed thin line) influenced by the finite-size effect. $n_{\text{app}}$ (dark green dashed/solid line) is further biased downward from $n'_{\text{app}}$ due to censorship related to the cut-off magnitude ($M_{\text{co}}$), which is empirically and conservatively set slightly above the completeness magnitude ($M_{\text{c}}$), with its standard deviation (dark green dashed/solid thin line) also affected by the finite-size effect. Recall that the smallest magnitudes of earthquakes that can be triggered and can trigger other earthquakes are represented by $m_*$ and $m_0$, as defined in Formulas (2) and (3) respectively. In order to simplify model calibration, it is the standard practice to assume that $m_*$ is equal to $m_0$.

All synthetic catalogs exhibit systematic branching ratio bias processes, which are summarised in Figure 1. Broadly, the biases affecting the branching ratio can be divided into three categories: (1) spatial and temporal boundary effects from catalog selection lead to a downward bias from the theoretical true branching ratio $n_{\text{true}}$ to the prime apparent branching ratio $n'_{\text{app}}$; (2) the existence of a cut-off magnitude $M_{\text{co}}$ for calibrating models introduces a further downward bias from $n'_{\text{app}}$ to $n_{\text{app}}(M_{\text{co}})$; (3) finite-size effects due to finite earthquake sample sizes result in statistical fluctuations of the estimated $n'_{\text{app}}$ and $n_{\text{app}}$ whose standard deviations are all the larger, the smaller the earthquake datasets.

The properties of the biases described in Figure 1 allow us to propose a correction method that begins with $n_{\text{app}}(M_{\text{co}})$, progresses to $n'_{\text{app}}$, and finally yields a refined estimate for $n_{\text{true}}$. This correction method exploits the information contained in the function $n_{\text{app}}(M_{\text{co}})$, which is much richer than the point estimate of $n_{\text{app}}$ at $M_{\text{co}}^{\text{best}}$. Our correction method starts by constructing the function $n_{\text{app}}(M_{\text{co}})$ that quantifies the impact of censorship. Through observation and testing of simulated data, we propose to fit $n_{\text{app}}(M_{\text{co}})$ by the model

$$n_{\text{app}}(M_{\text{co}}) = p_1 10^{p_2 M_{\text{co}}} + n'_{\text{app}} + x_1 \sigma_1 [N(m \geq M_{\text{co}})] \quad (15)$$

where $p_1 < 0$ and $p_2 > 0$ are two scalar parameters and the intermediate corrected branching ratio $n'_{app}$ ($> n_{app}(M_{co})$) is obtained as a calibrated parameter, together with $p_1$ and $p_2$. Accounting for the finite-size effects, $\sigma_1[N(m \geq M_{co})$ is the standard deviation of $n_{app}(M_{co})$ associated with the finite number of earthquakes of magnitudes larger than $M_{co}$. $x_1$ is a random variable with zero mean and unit variance, expressing that $n_{app}(M_{co})$ contains a stochastic component. For a given function $n_{app}(M_{co})$ that depends on the geometry of the auxiliary spatial domains, on the auxiliary temporal bands and on the other seismological parameters specific to the studied catalogue, our procedure consists in fitting Formula (15) to the function $n_{app}(M_{co})$ obtained by calibrating the ETAS model and thus obtain the partially corrected $n'_{app}$. The model quantifies how $n_{app}$ moves away from $n'_{app}$ as $M_{co}$ is increased. Inverting this relationship provides an estimate correction of $n_{app}(M_{co})$ into $n'_{app}$.

The last step consists in identifying the relationship between $n_{true}$ and $n'_{app}$. Our simulation experiments indicate that this relationship is approximately linear when $n'_{app} < 1$, which leads to our second proposed correction

$$n_{true} = q_1 n'_{app} + q_2 + x_2\sigma_2[N(m \geq m_0)] \tag{16}$$

where $q_1$ and $q_2$ are two positive scalar parameters. Accounting for the finite-size effects, $\sigma_2[N(m \geq m_0)]$ is the standard deviation of $n'_{app}$ associated with the finite number of earthquakes of magnitudes larger than $m_0$. $x_2$ is a random variable with zero mean and unit variance, expressing that $n_{true}$ contains a stochastic component. Our method thus establishes the relationship between $n_{app}$ and $n_{true}$. We will apply this bias correction framework to three observed catalogs. Upon obtaining $n_{true}$ from the estimated $n_{app}$, as a consistency check, we generate again synthetic catalogs with this $n_{true}$ estimate and reapply the bias correction framework to these catalogs to validate that our correction from $n_{app}(M_{co})$ for the real catalog to $n_{true}$ is accurate, as elaborated in Section 5.4.

## 4 Dataset

The earthquake catalogs utilized in the present study encompass three distinct regions: California, New Zealand, and the China Seismic Experimental Site (CSES) in Sichuan and Yunnan provinces (Wu & Li, 2021a; 2021b; Wu, 2022). The dataset spans from January 1, 1970 to October 27, 2023, for California, January 1, 1970 to December 12, 2023, for New Zealand, and January 1, 1970 to August 23, 2023, for CSES. For each region, earthquakes with depths down to 100 km and with $M \geq 0$ are included, as shown in Figures S1 to S3. Figures S1 to S3 also present the statistical characteristics of these earthquake catalogs. The completeness magnitude ($M_c$) is calculated using a one-year moving window with a two-year step. The method for quantifying power-law behavior of the complementary cumulative frequency-magnitude distribution, as proposed by Clauset et al. (2009), is employed for this calculation. Additionally, the variation of the $b$-value with $M_{co}$ is explored, determined through the maximum likelihood method proposed by Aki (1965). The ± 1σ standard deviation of the $b$-value is obtained through bootstrapping (Efron, 1979). Based on the results, the primary catalog supplied to the ETAS model starts in 1985 for California, in 1990 for New Zealand, and in 2000 for the CSES, each with an auxiliary time band of 15, 20, and 30 years, respectively (Table S1; Figures S1 to S3). The primary area for each region is outlined, accompanied by an auxiliary band extending approximately 100 km outward (Table S1; Figures S1 to S3).

## 5. Results

5.1 Estimation of the apparent branching ratio $n_{app}(M_{co})$

Figure 2 shows how $n_{app}$ varies as a function of $M_{co}$ using different estimation methods. It also tests the consistency of various $n$-estimation methods under four distinct scenarios (ordered from top to bottom):

-Scenario 1 involves the primary catalog with an auxiliary spatio-temporal band;

-Scenario 2 pairs the primary catalog with an auxiliary spatial band only;

-Scenario 3 uses the primary catalog with an auxiliary temporal band only; and

-Scenario 4 combines the primary catalog with the catalog from the auxiliary region, incorporating an auxiliary temporal band.

The examined estimation methods include the formulaic approach utilizing Formulas (6) and (7), the counting approach based on Formula (8), and the mean-variance-based estimation approach in Formula (12).

Figure 2 reveals that the mean-variance-based estimation approach, as per Formula (12), yields the largest estimates for $n$, approaching $n_c = 1$. However, these estimations of the branching ratio should not be trusted when dealing with spatially non-uniform background rates in real world catalogs, as noted by Wheatley et al. (2019) and Wehrli et al. (2021). Conversely, the formulaic approach Formula (7) results in the smallest $n$ estimates. Both methods exhibit significant deviations from the estimates produced by the formulaic approach Formula (6) and the counting approach Formula (8). The latter two methods give estimates of comparable magnitudes and display trends in the variation of $n_{app}$ with $M_{co}$ that are similar to those observed in the synthetic experiments illustrated in Figure 1. The estimates from the counting approach, governed by Formula (8), exhibit a more stable curve with clearer trends compared with that of the formulaic approach Formula (6), primarily due to the reliance of the latter on the need for accurately estimated parameters $b$, $K$, and $a$, along with reasonable assumptions for $m_0$ and $m_{max}$. The accuracy of the counting approach hinges on precise estimates of $N_{tri}$ or $N_{bkg}$.

It is useful to compare with the results of Nandan et al. (2021b), who estimated the branching ratios for California and New Zealand to be 0.79 and 0.61, respectively, using Formula (6). They used the primary catalog for both regions covering the period from 1981 to 2020 (approximately 40 years), with $M_{co}$ set at 3.0 and 4.0 for each region, respectively. Their set-up corresponds to the Scenario 3 shown as the third row of panels in Figure 2. Scenario 3 only takes into account an auxiliary temporal band in the calibration of the ETAS model. The spatial extent of the primary catalog for California in the present study is exactly the same as that of Nandan et al. (2021b), and the time range is also similar (Table S1). Therefore, the branching ratio ($n = 0.79$) estimated using Formula (6) in Scenario 3 of Figure 2 in the present study, with $M_{co} = 3.5$, is basically consistent with the results of Nandan et al. (2021b). However, Figure 2 also shows that using different auxiliary catalogs can lead to significant differences. For example, considering only an auxiliary spatial band results in an estimated branching ratio of about 0.6; Scenario 1, which is considered a priori to be the most complete and involves an auxiliary spatio-temporal band, yields an estimated branching ratio of about 0.7. For New Zealand, there are significant differences between the spatio-temporal range of the primary catalog used in the present study and that of Nandan et al. (2021b). As a result, the branching ratio estimated using

Formula (6) in Scenario 3 of Figure 2 for $M_{co} = 4.0$ is about 0.79, whereas the estimate by Nandan et al. (2021b) is 0.61. In the following, we will revisit seismicity criticality in the light casted by our understanding of the three biases and how their corrections will impact the results.

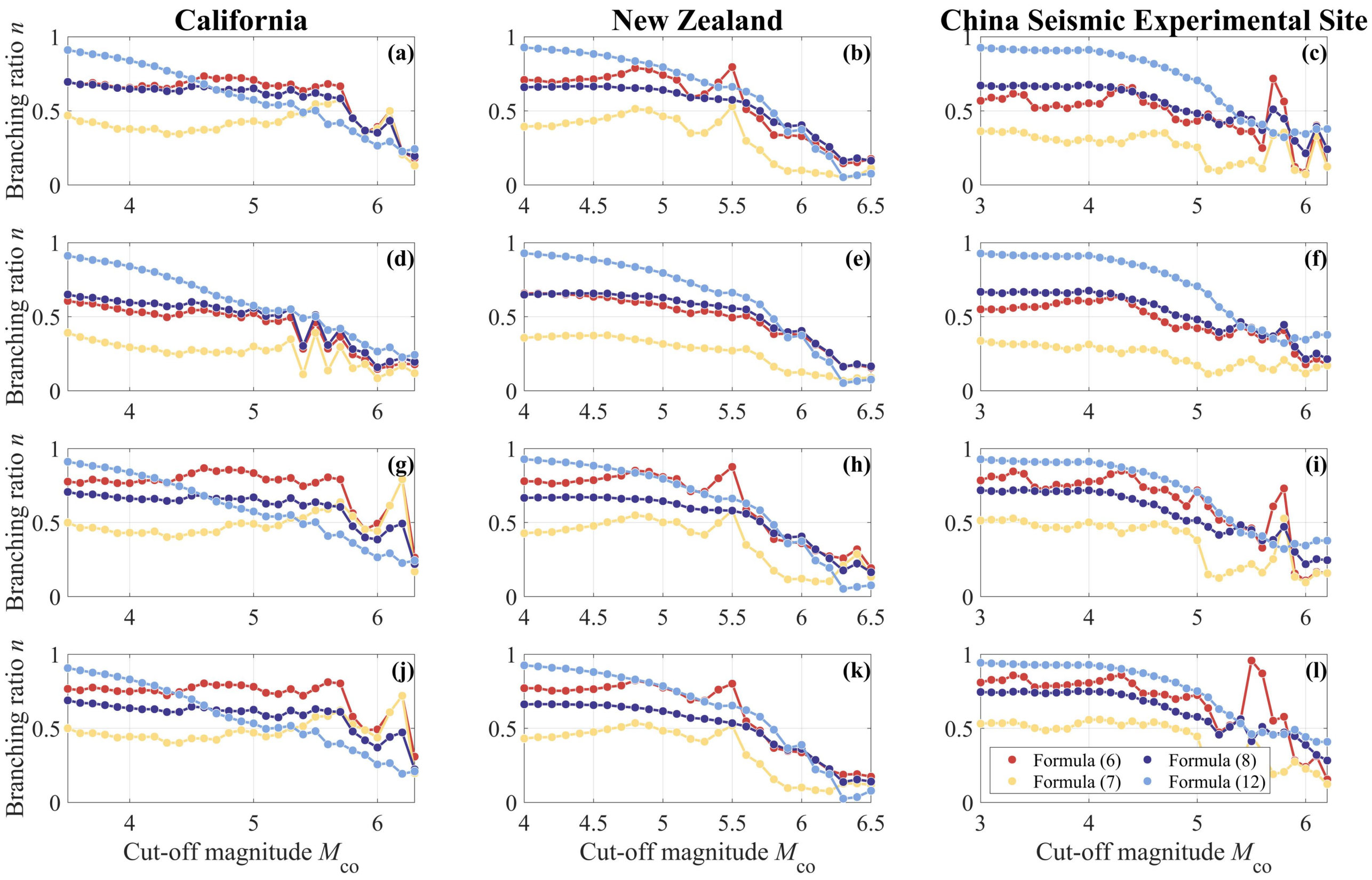

**Figure 2**. Estimation of the apparent branching ratio $n_{app}$ using multiple methods, shown as a function of the cut-off magnitude $M_{co}$ for California, New Zealand, and CSES (arranged from left to right). The panels, ordered from top to bottom, represent: (1) The primary catalog with an auxiliary spatio-temporal band; (2) The primary catalog with an auxiliary spatial band only; (3) The primary catalog with an auxiliary temporal band only; (4) The primary catalog combined with the catalog from the auxiliary region, with an auxiliary temporal band.

### 5.2 Censorship correction: From $n_{app}(M_{co})$ to $n'_{app}$

The censorship correction proposed in the present study relies on the information provided by the dependence of the apparent estimated branching ratio ($n_{app}$) as a function of different cut-off magnitudes ($M_{co} \geq M_{co}^{best}$) as depicted in Figure 2. We use Formula (15) that has been motivated and validated on synthetic catalogues as explained in Section 3. Formula (15) describes the variation of $n_{app}$ as a function of $M_{co}$, and has three adjustable parameters $\{p_1, p_2, n'_{app}\}$. For each earthquake catalogue, we calibrate the ETAS model for different value of $M_{co}$ and obtain the function $n_{app}(M_{co})$ that is specific to the geometry of the auxiliary spatial domains, to the temporal band and to the other seismological parameters specific to the studied catalogue. We then fit Formula (15) via a standard nonlinear least-square fitting procedure to the function $n_{app}(M_{co})$ determined from the ETAS calibration to obtain the partially corrected branching ratio $n'_{app}$. In the fitting procedure of Formula (15), we give more weights to the value of $M_{co}$ closest to $M_{co}^{best}$ as they correspond to catalogues with larger numbers of earthquakes. Typically, for each $M_{co}$ value, one could repeatedly recalibrate the ETAS model to obtain the estimated mean and standard deviation of $n_{app}$. However, during the recalibration process of the ETAS model, we observed that the standard deviation of $n_{app}$ across recalibrations was consistently minor, fluctuating with an order of magnitude of ±0.0001. Given such negligible variations quantified by a small standard deviation and in the interest of conserving computational resources, we opted not to estimate the standard deviation of $n_{app}$ in our study. Instead, we assumed the standard deviation to be zero, that is, $\sigma_1[N(m \geq M_{co})] = 0$.

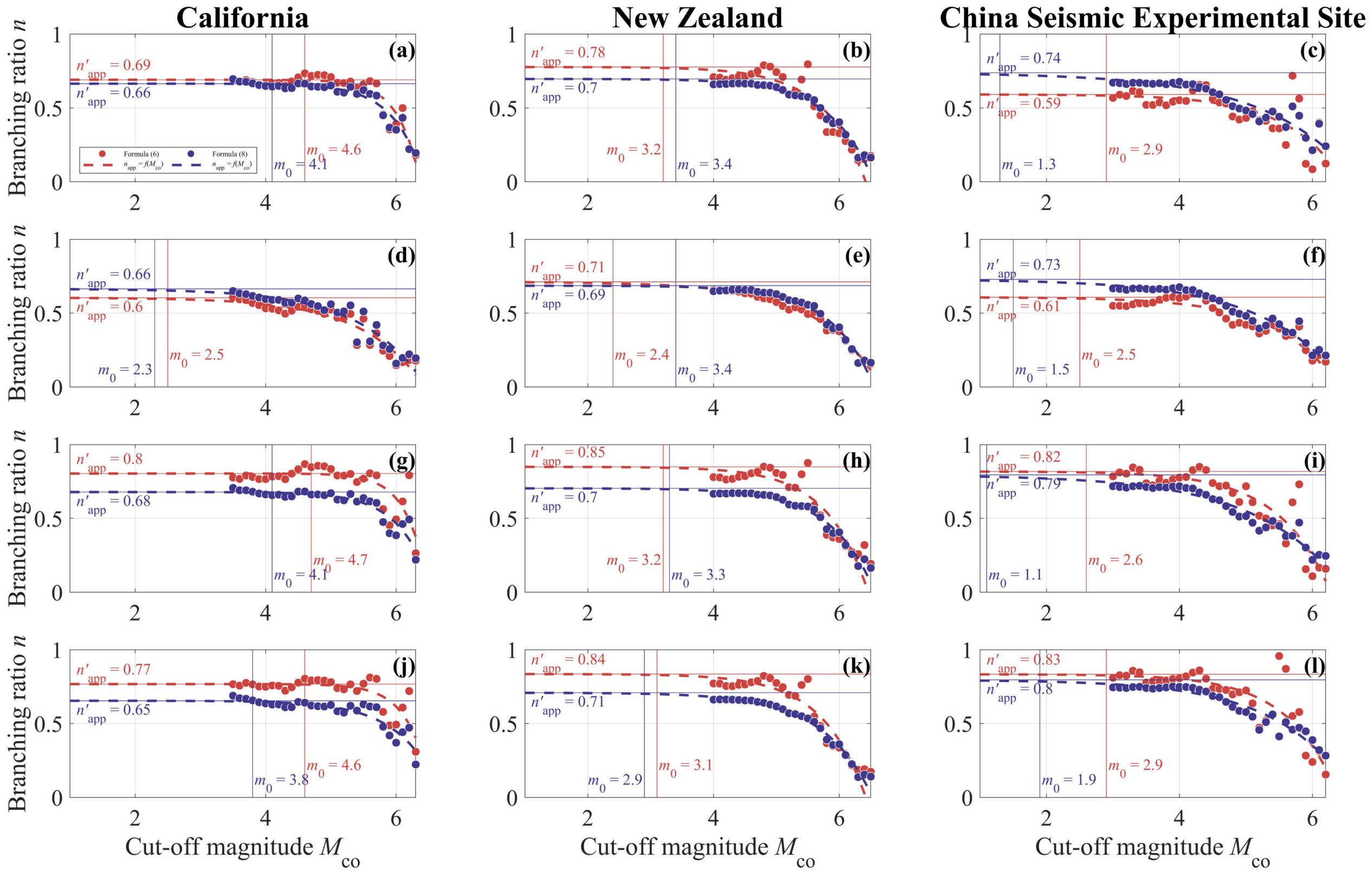

**Figure 3**. Correction of censorship effects using model given by Formula (15) from which the partial corrected branching ratio $n'_{app}$ is obtained from $n_{app}$ for California (left column), New Zealand (middle column), and CSES (right column). The four rows of panels corresponds to the same four scenarios as shown in Figure 2. The value of $M_{co}$ at which the model's curve begins to level off is identified as $m_0$ as explained in Section 5.2. As in Figure 2, the red data points are obtained with Formula (6) and the dark blue data points are obtained with Formula (8).

Figure 3 shows the fits of $n_{app}(M_{co})$ by Formula (15) for the three regions of California, New Zealand and CSES, for the four scenarios and for the two calculation methods given by Formulas (6) and (8) for the branching ratio. Table 1 gives the fitted model parameter values for $n_{app}(M_{co})$ determined by Formula (6). Table 2 gives the fitted model parameter values for $n_{app}(M_{co})$ determined by Formula (8). Except for Scenario 2 in California and Scenarios 1 and 2 in CSES, Formula (6) consistently yields larger branching ratio estimates than Formula (8) for all the other scenarios (Figure 3; Tables 1 and 2). This is primarily because the branching ratios estimated based on Formula (6) correct for the boundary effect to some extent during parameter estimation, compared to those based on Formula (8). The branching ratios estimated by Formula (6) show more pronounced variations with changes in $M_{co}$, especially at larger $M_{co}$.

**Table 1.** Parameters of Formula (15) $n_{app} = p_1 10^{p_2 M_{co}} + n'_{app}$ fitted to the function $n_{app}(M_{co})$ determined from the ETAS calibrations with Formula (6) of the three regions (columns) for the four scenarios (rows).

| | California | | | | New Zealand | | | | CSES | | | |
|---|---|---|---|---|---|---|---|---|---|---|---|---|
| | $m_0$ | $p_1$ | $p_2$ | $n'_{app}$ | $m_0$ | $p_1$ | $p_2$ | $n'_{app}$ | $m_0$ | $p_1$ | $p_2$ | $n'_{app}$ |
| **Scenario 1** | 4.6 | $-3.12\times10^{-9}$ | 1.31 | 0.69 | 3.2 | $-5.35\times10^{-5}$ | 0.65 | 0.78 | 2.9 | $-2.43\times10^{-4}$ | 0.52 | 0.59 |
| **Scenario 2** | 2.5 | $-6.00\times10^{-4}$ | 0.47 | 0.60 | 2.4 | $-8.72\times10^{-4}$ | 0.44 | 0.71 | 2.5 | $-5.98\times10^{-4}$ | 0.47 | 0.61 |
| **Scenario 3** | 4.6 | $-1.34\times10^{-9}$ | 1.35 | 0.80 | 3.2 | $-4.15\times10^{-5}$ | 0.67 | 0.85 | 2.6 | $-2.46\times10^{-4}$ | 0.56 | 0.82 |
| **Scenario 4** | 4.6 | $-4.10\times10^{-9}$ | 1.26 | 0.77 | 3.1 | $-6.56\times10^{-5}$ | 0.64 | 0.84 | 2.9 | $-1.40\times10^{-4}$ | 0.59 | 0.83 |

**Table 2.** Parameters of Formula (15) $n_{app} = p_1 10^{p_2 M_{co}} + n'_{app}$ fitted to the function $n_{app}(M_{co})$ determined from the ETAS calibrations with Formula (8) of the three regions (columns) for the four scenarios (rows).

| | California | | | | New Zealand | | | | CSES | | | |
|---|---|---|---|---|---|---|---|---|---|---|---|---|
| | $m_0$ | $p_1$ | $p_2$ | $n'_{app}$ | $m_0$ | $p_1$ | $p_2$ | $n'_{app}$ | $m_0$ | $p_1$ | $p_2$ | $n'_{app}$ |
| **Scenario 1** | 4.1 | $-5.08\times10^{-7}$ | 0.95 | 0.66 | 3.4 | $-4.61\times10^{-5}$ | 0.64 | 0.70 | 1.3 | $-4.84\times10^{-3}$ | 0.33 | 0.74 |
| **Scenario 2** | 2.3 | $-8.79\times10^{-4}$ | 0.44 | 0.66 | 3.4 | $-4.46\times10^{-5}$ | 0.64 | 0.69 | 1.5 | $-3.44\times10^{-3}$ | 0.36 | 0.73 |
| **Scenario 3** | 4.1 | $-8.15\times10^{-7}$ | 0.90 | 0.68 | 3.3 | $-5.66\times10^{-5}$ | 0.62 | 0.70 | 1.1 | $-5.23\times10^{-3}$ | 0.33 | 0.79 |
| **Scenario 4** | 3.8 | $-1.19\times10^{-5}$ | 0.71 | 0.65 | 2.9 | $-1.84\times10^{-4}$ | 0.55 | 0.71 | 1.0 | $-2.01\times10^{-3}$ | 0.39 | 0.80 |

Examining Scenarios 1 and 3 for the three study regions, which involve the same primary catalogs with varying auxiliary spatial bands, we observe that Formula (6) leads to approximately a +0.1 to +0.2 bias in the $n$ estimates. In contrast, when inspecting the scenarios with different auxiliary temporal bands (Scenarios 1 and 2), the influence on Formula (6) is smaller. For California and New Zealand, the impact is around -0.08, while for CSES, it is approximately +0.02. Formula (8) shows minimal deviation (-0.01 to +0.05) in estimates across different scenarios. The deviations in $n'_{app}$ fitted from $n_{app}(M_{co})$ obtained using Formula (6) are larger for the different scenarios in the three study regions compared to the results from Formula (8), with $n'_{app}$ values ranging from 0.60 to 0.80 for California, 0.71 to 0.85 for New Zealand, and 0.59 to 0.83 for CSES. $n'_{app}$ values obtained from Formula (8) range from 0.66 to 0.68 for California, 0.69 to 0.71 for New Zealand, and 0.73 to 0.80 for CSES. We hypothesize that this stability may be due to a compensatory effect at the spatiotemporal boundaries, where the number of triggered earthquakes classified as background earthquakes balances out with the number of background earthquakes classified as triggered. Therefore, this estimation method, which relies on the proportion of triggered earthquake counts, appears to be largely unaffected by variations in the auxiliary spatio-temporal band. Moreover, considering the need to synthesize a large number of simulated earthquake catalogs and calculate $n'_{app}$ for each catalogue in the next step, the computation of $n'_{app}$ based on Formula (6) requires calibrating each set of simulated catalogs using the ETAS model, demanding significant computational power and time. In contrast, Formula (8) simply requires calculating the proportion of triggered earthquakes in the simulated catalogs, which is much simpler. Therefore, the present study will proceed with corrections using $n'_{app}$ from Formula (8) for further analysis.

The censorship correction model also provides information about $m_0$, the smallest magnitude of earthquakes capable of triggering other earthquakes defined in Formula (3). We propose that the largest $M_{co}$ below which the curve of $n_{app}$ starts to flatten as $M_{co}$ decreases corresponds to $m_0$, as sketched in Figure 1. The reasoning is that including smaller earthquakes in the ETAS calibration does not alter the branching ratio, indicating that we accurately account for all triggering earthquakes above this characteristic magnitude. It is determined in the present study as the minimum $M_{co}$ at which the derivative of the censorship correction formula $n_{app} = p_1 10^{p_2 M_{co}} + n'_{app}$ is larger than the threshold -0.01. Figure 3 and Tables 1 and 2 present the estimated $m_0$ based on Formulas (6) and (8). Taking the average of the two $m_0$ estimates gives $m_0$ values of 4.4, 2.4, 4.4, 4.2 for the four scenarios in California; 3.3, 2.9, 3.3, 3.0 for the four scenarios in New Zealand; and 2.1, 2.0, 1.9, 2.4 for the four scenarios in CSES, respectively.

### 5.3 Boundary and finite-size effects correction: From $n'_{app}$ to $n_{true}$

As discussed in the introduction, the boundary effect is related to the temporal and spatial extent of the study region, while the finite size effect results from the finite size of the seismicity sample. We propose to correct $n'_{app}$ in order to recover $n_{true}$ by synthesizing catalogs with the same spatiotemporal range. Specifically, we synthesize earthquake catalogs using calibrated ETAS model parameter sets $\{Q, D, c, \omega, \tau, d, \gamma, \rho\}$ at $M_{co}^{best}$ (Table S2), combined with the aforementioned estimation of $m_0$. Since the triggering relationships among earthquakes in the simulated catalogs are known, it is straightforward to calculate $n'_{app}$ by using Formula (8).

Figure 4 shows the dependence of $n'_{app}$ and its standard deviation as a function of $n_{true}$ quantifying the impact of boundary effects in time and space, with standard deviation influenced by the finite-size effects, for California, New Zealand, and CSES. The mean and standard deviation of $n'_{app}$ are calculated from 100 simulated earthquake catalogs for each $n_{true}$, for the three study regions under different scenarios. In Section 3, we have established through numerical simulations the relationship between $n_{true}$ and $n'_{app}$ given by Formula (16) for $n_{true} < 1$. The simulation experiments provide the standard deviation of $n'_{app}$, denoted as $\sigma_2[N(m \geq m_0)]$. According to error propagation theory, we use the following formula to estimate the standard deviation of $n_{true}$ based on $\sigma_2[N(m \geq m_0)]$, that is,

$$\sigma_{n_{\text{true}}} \approx |\frac{\mathrm{d}(q_1 n'_{\text{app}} + q_2)}{\mathrm{d}n'_{\text{app}}}| \cdot \sigma_2[N(m \geq m_0)] = q_1 \cdot \sigma_2[N(m \geq m_0)] \tag{17}$$

It should be noted that $\sigma_{true}$ is essentially an estimate derived from $n'_{app}$ specifically for correcting boundary effects, not encompassing the entire correction process from $n_{app}$. In fact, $\sigma_{true}$ should integrate both boundary effects and censorship correction uncertainties. However, since we do not delve into the uncertainties of censorship correction, implicitly setting $\sigma_1[N(m \geq M_{co})] = 0$, $\sigma_{true}$ is exclusively associated with uncertainties related to boundary effects.

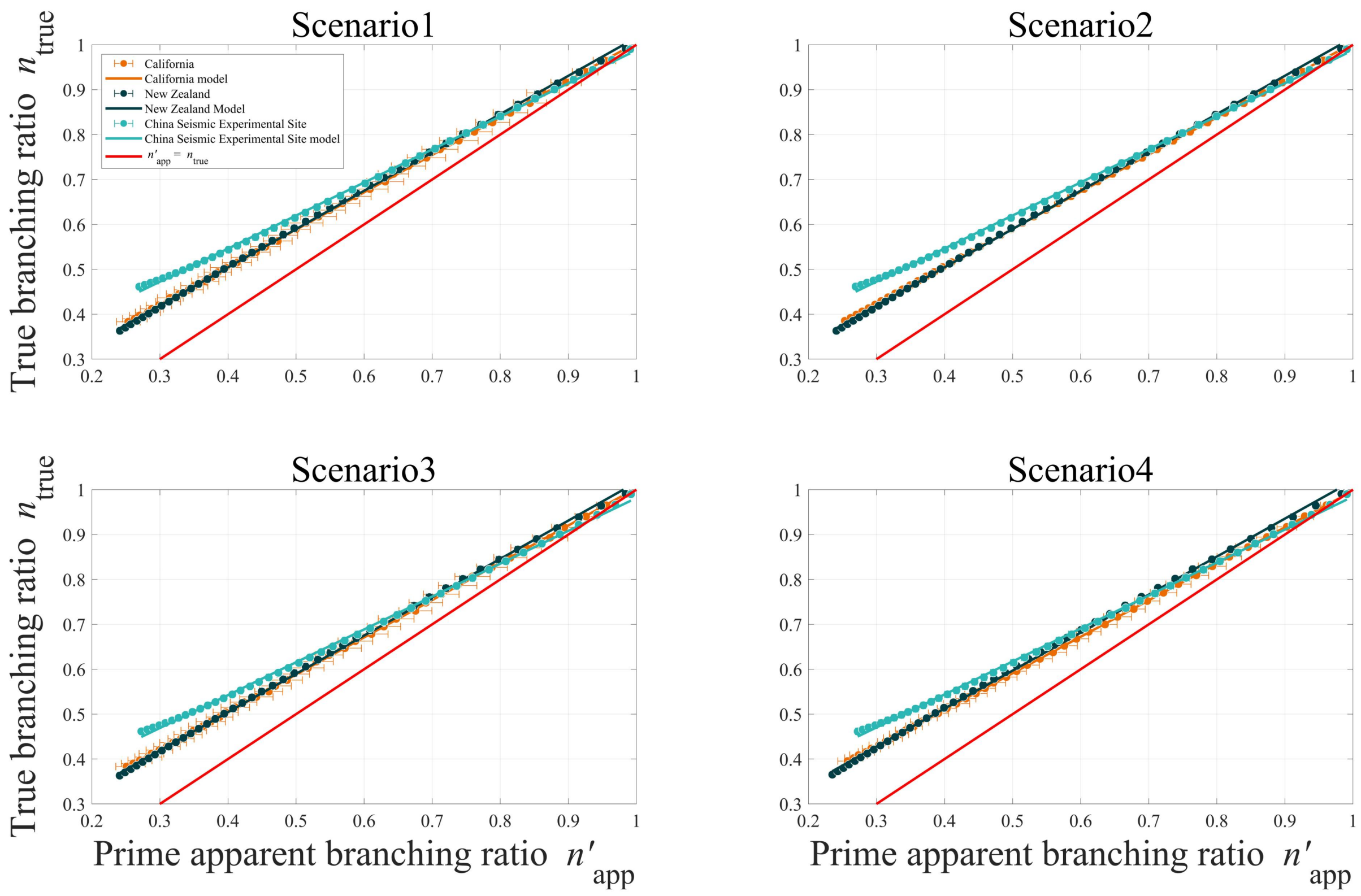

**Figure 4**. Dependence of $n'_{app}$ and its standard deviation as a function of $n_{true}$ (in reverse axis) showing the quantitative impact of boundary effects in time and space, with standard deviation influenced by the finite-size effect, for California, New Zealand, and CSES. For each value of $n_{true}$, 100 sets of catalogs have been simulated, mirroring the time and space ranges of the primary catalog in the three study regions. The mean and variance of $n'_{app}$ over these 100 sets of simulated catalogs are calculated using the counting approach given by Formula (8). The observed linear dependence suggests the correction model given by $n_{true} = q_1 n'_{app} + q_2$ which allows us to extract $n_{true}$ from $n'_{app}$.

The solid lines in Figure 4 are the best fits with Formula (16) to $n'_{app}(n_{true})$ for different scenarios in each study region. Table 3 lists the values of the parameters of Formula (16) and the corresponding final estimates of $n_{true}$. Accounting for boundary effects and finite-size effects, the initial estimated $n_{app}$ needs to be corrected upwards by approximately 0.04 to 0.06 to obtain $n_{true}$. The final $n_{true}$ estimates are 0.71 to 0.74 for California, 0.75 to 0.77 for New Zealand, and 0.79 to 0.84 for CSES. The seismicity within these regions is closer to criticality than inferred without corrections, yet still maintains a certain distance from the critical value $n_c = 1$.

Compared to the estimates by Nandan et al., (2021b) for the global ($n = 0.45$), California ($n = 0.79$), and New Zealand ($n = 0.61$), our results feature a smaller standard deviation across different scenarios and regions. The results of Nandan et al. (2021b) for the global catalog were significantly influenced by censorship due to the setting of $M_{co} = 5$. Meanwhile, their estimates for California and New Zealand did not account for boundary effects in time and space and for the finite size effects, thus introducing some bias and uncertainty in their point estimates for these two regions. Our current findings imply that about 70% to 85% of observed seismicity is triggered by preceding earthquakes, with the remaining 15% to 30% being background seismicity driven by the forces of plate tectonics.

**Table 3.** Parameters for the model $n_{true} = q_1 n'_{app} + q_2$, utilized to correct boundary effects in time and space, where standard deviation is impacted by the finite-size effect. In this table, $n'_{app}$ is derived by fitting the $n_{app}$ values estimated using Formula (5), as shown in Table 3.

| | California | | | | New Zealand | | | | CSES | | | |
|---|---|---|---|---|---|---|---|---|---|---|---|---|
| **Scenario 1** | $n'_{app}$ | $q_1$ | $q_2$ | $n_{true} \pm \sigma$ | $n'_{app}$ | $q_1$ | $q_2$ | $n_{true} \pm \sigma$ | $n'_{app}$ | $q_1$ | $q_2$ | $n_{true} \pm \sigma$ |
| | 0.66 | 0.83 | 0.17 | 0.72 ± 0.0211 | 0.70 | 0.85 | 0.16 | 0.76 ± 0.0036 | 0.74 | 0.74 | 0.25 | 0.80 ± 0.0027 |
| **Scenario 2** | $n'_{app}$ | $q_1$ | $q_2$ | $n_{true}$ | $n'_{app}$ | $q_1$ | $q_2$ | $n_{true}$ | $n'_{app}$ | $q_1$ | $q_2$ | $n_{true}$ |
| | 0.66 | 0.83 | 0.18 | 0.72 ± 0.0023 | 0.69 | 0.85 | 0.16 | 0.75 ± 0.0025 | 0.73 | 0.74 | 0.25 | 0.79 ± 0.0025 |
| **Scenario 3** | $n'_{app}$ | $q_1$ | $q_2$ | $n_{true}$ | $n'_{app}$ | $q_1$ | $q_2$ | $n_{true}$ | $n'_{app}$ | $q_1$ | $q_2$ | $n_{true}$ |
| | 0.68 | 0.83 | 0.17 | 0.74 ± 0.0202 | 0.70 | 0.85 | 0.16 | 0.76 ± 0.0039 | 0.79 | 0.73 | 0.25 | 0.83 ± 0.0022 |
| **Scenario 4** | $n'_{app}$ | $q_1$ | $q_2$ | $n_{true}$ | $n'_{app}$ | $q_1$ | $q_2$ | $n_{true}$ | $n'_{app}$ | $q_1$ | $q_2$ | $n_{true}$ |
| | 0.65 | 0.81 | 0.18 | 0.71 ± 0.0166 | 0.71 | 0.85 | 0.17 | 0.77 ± 0.0023 | 0.80 | 0.73 | 0.25 | 0.84 ± 0.0031 |

## 6. Discussion and conclusions

We have proposed a new method to correct the three major identified biases by utilizing the functional dependence of estimated ETAS parameters on the artificial increase of the cut-off magnitude $M_{co}$. Our findings reveal that by accurately considering not just the spatial variation of the background rate but also censorship, temporal and spatial boundary effects, and the finite-size effects of the earthquake catalog, we obtain a branching ratio of seismicity that is closer to criticality than inferred without corrections, yet still maintains a certain distance from it, with values ranging from 0.70 to 0.85. It is interesting that, $m_0$ estimated by our methods is found as large as around 4 for California, 3 for New Zealand and 2 for CSES.

The minimum magnitude $m_0$ for earthquake triggering capability, an intrinsic ingredient of the ETAS model, and the branching ratio *n* demonstrate significant variability for different tectonic regimes, with California displaying the largest $m_0$, followed by New Zealand, and CSES presenting the lowest. Conversely, the branching ratio *n* is observed to exhibit an inverse relationship: it is largest in CSES, intermediate in New Zealand, and lowest in California.

Let us propose a tectonic interpretation of these results. The mechanism of horizontal sliding along California's transform faults, typically associated with efficient energy dissipation, tends to minimize aftershock occurrences (Kanamori & Brodsky, 2004; Stein & Wysession, 2009). Conversely, the vertical plate movements characterizing New Zealand's subduction zones introduce activations of seismicity at intermediate depths, complicating rupture dynamics and potentially amplifying aftershock activity due to the interaction of seismic waves with varied geological structures (Lay & Wallace, 1995; Scholz, 2019). In CSES, the intense crustal stresses arising from the collision between the Indian and Eurasian plates engender a complex network of faults. This complexity, coupled with the shallow depth of seismicity, facilitates the activation of numerous faults, thereby enhancing the region's aftershock triggering capability (China Earthquake Administration, 2019). Moreover, fault roughness, by influencing co-seismic slip, also modulates aftershock patterns, underscoring the intricate relationship between geological structures and types of seismicity (e.g., Cochran et al., 2023; Goebel et al., 2023). Relating tectonic dynamics and fault morphology to characteristic properties of seismicity such as *n* and $m_0$ offers novel insights into regional seismicity pattern, emphasizing the potential crucial role of underlying geological conditions in modulating aftershock sequences. It is important to note that the issue with $m_0$ occurs in several ETAS models. However, this is not the case for other models which, for example, incorporate two branches of the Gutenberg-Richter law, like those suggested by Vere-Jones (2005), analyzed by Saichev & Sornette (2005), and more recently observed in EM stochastic reconstructions by Nandan et al. (2022).

The non-critical but still rather large values of the branching ratio $n$ prompt a reevaluation of our understanding of the brittle fracture process within the Earth's crust. So far, reported values of $n$ found close to $n_c = 1$ have aligned with the popular concept of self-organized criticality, suggesting that the loaded fault network is in a permanent critical state. In contrast, values significantly below $n_c = 1$ indicate that fault networks primarily evolve far from a critical point. Our estimates of $n$, derived from more appropriate assumptions and an improved research methodology, fall between these two extremes, neither continuously critical nor far from it. In fact, our results align with the current state of research on earthquake predictability, suggesting that earthquakes are not entirely unpredictable as they would be if the seismogenic crust was in a self-organized critical state (where any event is mechanistically indistinguishable from others, making it impossible to effectively identify precursors before an event occurs) (Geller et al., 1997). The degree of earthquake predictability may be in part associated with sporadic emergence of singularities appearing through various mechanisms that could help signal the approach of catastrophic events.

Finally, in addition to the factors discussed in this study, the fidelity of the model to actual seismicity is equally important for accurately gauging criticality. For example, including the depth component in the model formulation can greatly improve model fitting (Guo et al., 2015a, 2018; Zhuang et al., 2019), which also yields a smaller branching ratio. Another factor is the rupture geometry of large earthquakes (Hainzl et al., 2008; Guo et al., 2015b, 2019, 2021). Ignoring the volume of the earthquake's rupture but regarding the focal zone as a point leads to a larger value of $K$ and a smaller value of $\alpha$ (Zhuang et al., 2019; Guo et al., 2021). Nevertheless, even the estimation for these improved versions of the ETAS model also suffers from the problems discussed in this study. We leave these issues for future research.

## Acknowledgments

The authors acknowledge Dr. Shyam Nandan for sharing an ETAS model code from which our present code is derived. This work is partially supported by the Guangdong Basic and Applied Basic Research Foundation (Grant No. 2024A1515011568), the National Natural Science Foundation of China (Grant no. U2039202 and U2039207), Shenzhen Science and Technology Innovation Commission (Grant no. GJHZ20210705141805017), and the Center for Computational Science and Engineering at Southern University of Science and Technology.

**Open Research**

The California catalog is sourced from the Advanced National Seismic System (ANSS) Comprehensive Earthquake Catalog (ComCat), accessible at https://earthquake.usgs.gov/data/comcat/ (last accessed: December 18, 2023). For the catalog in New Zealand, the data are obtained from the GeoNet Earthquake Catalog of New Zealand, accessible at https://quakesearch.geonet.org.nz/ (last accessed: December 18, 2023). The data for the China Seismic Experimental Site are acquired from the China Earthquake Networks Center (CENC) through the internal link provided by the Earthquake Cataloging System at China Earthquake Administration, available at http://10.5.160.18/console/index.action (last accessed: December 18, 2023), with a Digital Object Identifier (DOI) of 10.11998/SeisDmc/SN.

Supporting Information for

# Revisiting Seismicity Criticality: A New Framework for Bias Correction of Statistical Seismology Model Calibrations

Jiawei Li[1], Didier Sornette[1], Zhongliang Wu[1,2], Jiancang Zhuang[1,3], and Changsheng Jiang[4]

[1]Institute of Risk Analysis, Prediction and Management (Risks-X), Academy for Advanced Interdisciplinary Studies, Southern University of Science and Technology (SUSTech), Shenzhen, China.

[2]Institute of Earthquake Forecasting, China Earthquake Administration, Beijing, China.

[3]The Institute of Statistical Mathematics, Research Organization of Information and Systems, Tokyo, Japan.

[4]Institute of Geophysics, China Earthquake Administration, Beijing, China.

## Contents of this file

## Introduction

In the present study, we identify and quantify three sources of bias: (i) boundary effects, (ii) finite-size effects, and (iii) censorship, which cause errors in seismic analysis and predictions. By employing a variant of the ETAS model with variable spatial background rates, we propose a method to correct for these biases, focusing on the branching ratio $n$, a key indicator of earthquake triggering potential. We validate our method using synthetic earthquake catalogs, accurately recovering the true branching ratio $n_{true}$ after correcting biases with $n_{app}$. Additionally, our method introduces a refined estimation of the minimum triggering magnitude $m_0$, a crucial parameter in the ETAS model. Applying our framework to the earthquake catalogs of California, New Zealand, and the China Seismic Experimental Site (CSES), we revisit seismicity's criticality to enhance our comprehension of seismic patterns, aftershock predictability, and inform earthquake risk mitigation and management strategies. The Supporting Information accompanying this study provides an in-depth statistical characterization of seismicity in the three study regions, along with the calibrated parameters of the Epidemic-Type Aftershock Sequences (ETAS) model used to simulate synthetic catalogs.

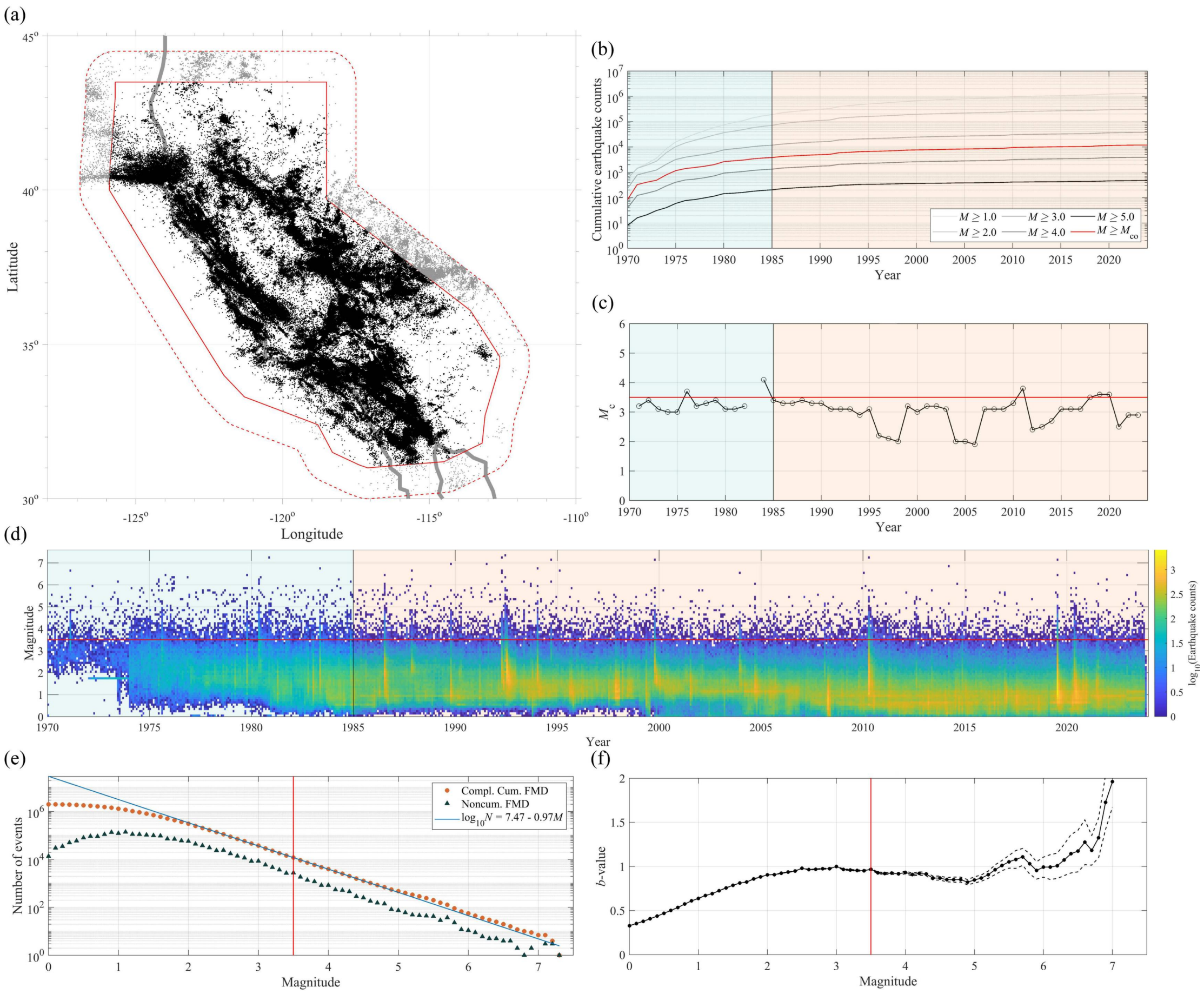

**Figure S1.** (a) Epicenter distribution of 1,994,000 recorded earthquakes from 1970 to 2023 in the primary California region (solid red line), complemented by an approximate 100 km auxiliary space band (dashed red line). The thick gray line delineates the coastline. (b) Time series depicting cumulative earthquake counts for different cut-off magnitudes ($M_{co} \geq 1$, $\geq 2$, $\geq 3$, $\geq 4$, and $\geq 5$). (c) Evolution of the completeness magnitude ($M_c$) over time. (d) Monthly earthquake counts per 0.1 magnitude bin. (e) Complementary cumulative and density frequency–magnitude distribution alongside the Gutenberg–Richter (GR) law fitted using earthquakes with magnitudes larger than $M_{co}^{best} = 3.5$. (f) Variation of the $b$-value with $M_{co}$, with the thin dashed line indicating the $\pm 1\sigma$ standard deviation. The red lines in panels (b) to (f) denote $M_{co}^{best} = 3.5$ in California. The light red and light blue shaded regions in panels (b) to (d) represent the primary period and the auxiliary time band, respectively.

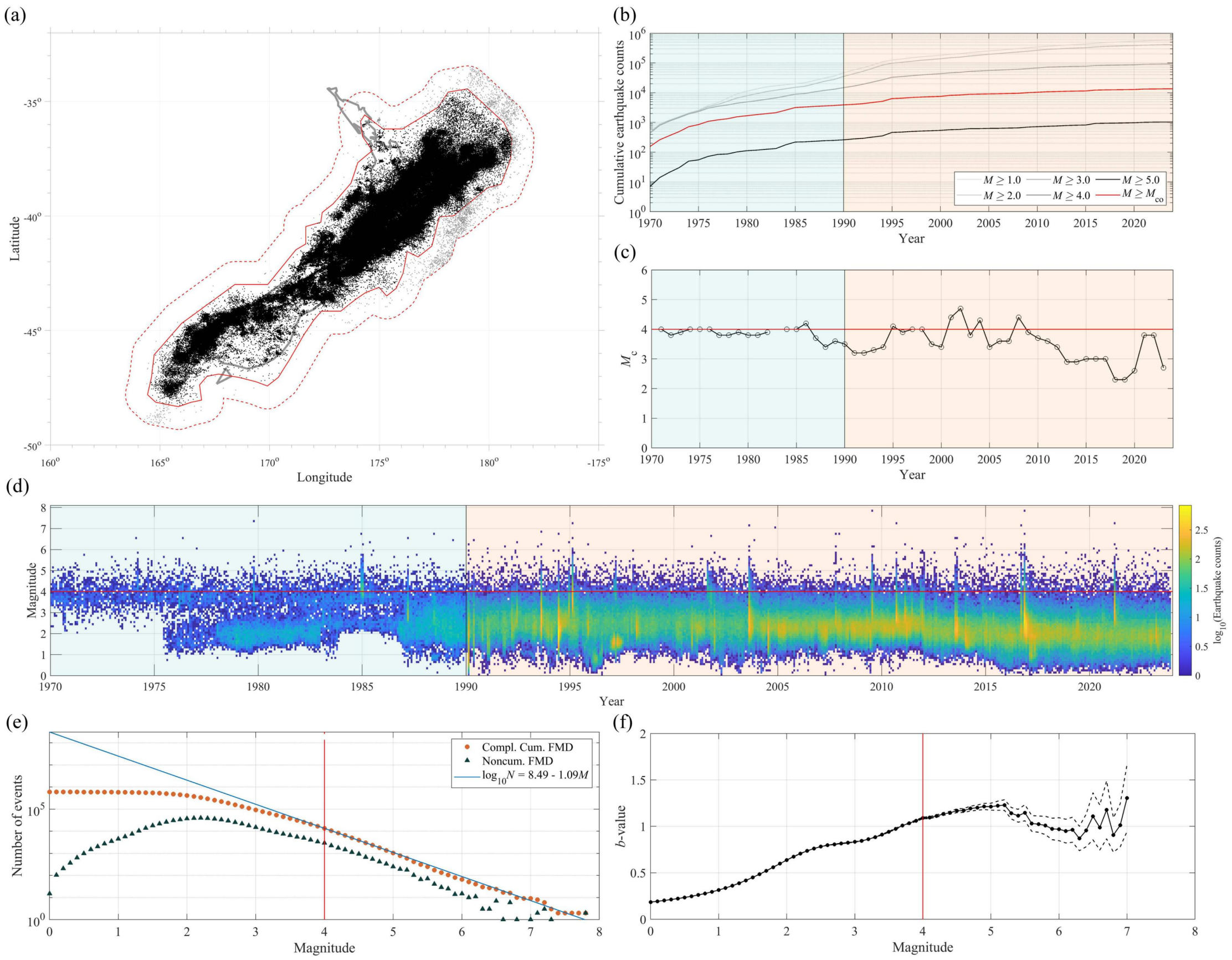

**Figure S2.** (a) Epicenter distribution of 593,000 recorded earthquakes from 1970 to 2023 in the primary New Zealand region (solid red line), complemented by an auxiliary space band (dashed red line). Panels (b) to (f) are the same as in Figure S1, with the red lines denoting $M_{co}^{best}$ = 4 in New Zealand.

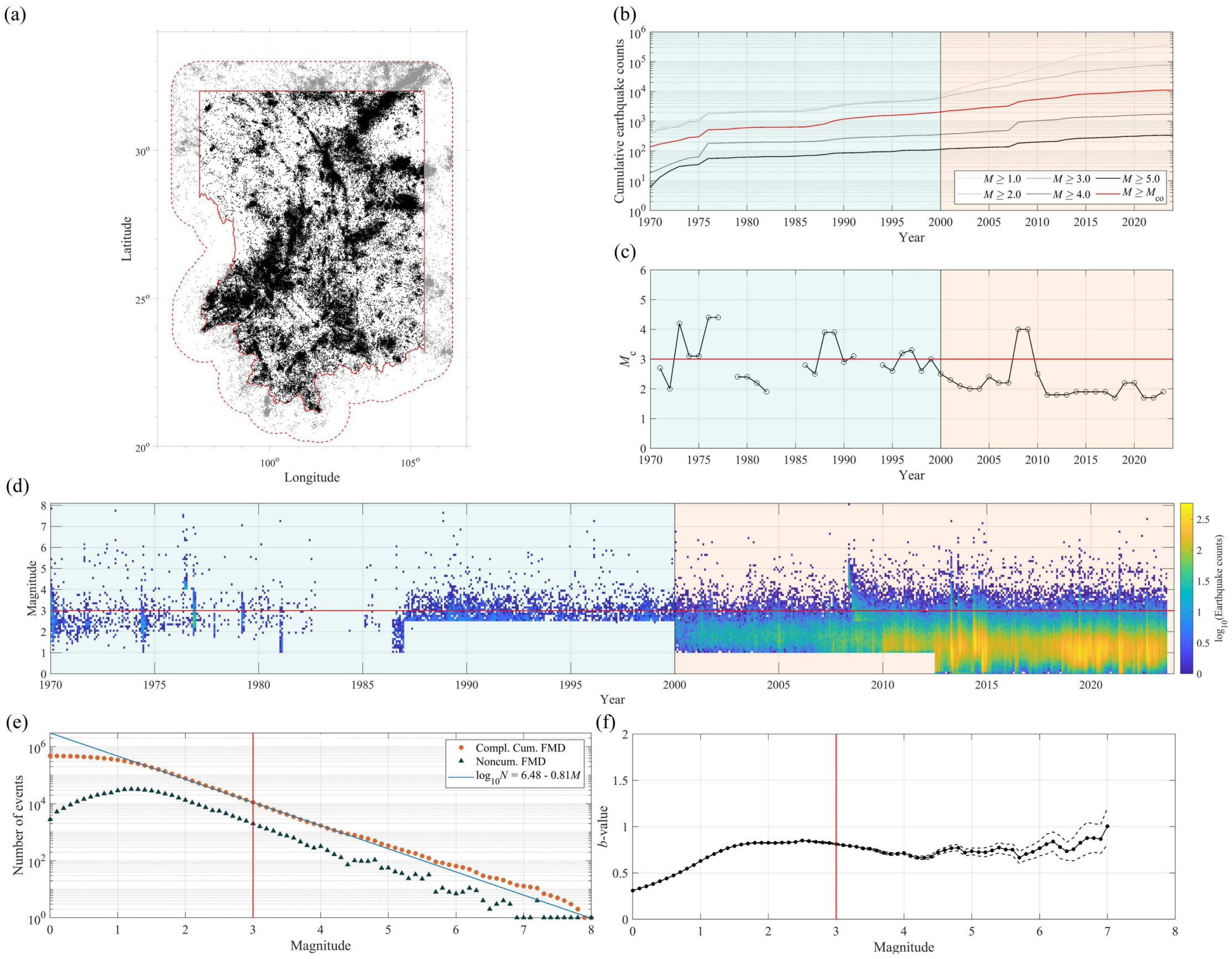

**Figure S3**. (a) Epicenter distribution of 477,000 recorded earthquakes from 1970 to 2023 in the primary region of the China Seismic Experimental Site (CSES; solid red line), complemented by an auxiliary space band (dashed red line). Panels (b) to (f) are the same as in Figures S1 and S2, with the red lines denoting $M_{co}^{best}$ = 3 in the Sichuan–Yunnan region, China.

**Table S1**. Main characteristics of the primary catalog used in this study.

| | Area [km$^2$] | Source | $M_{co}^{best}$ | Primary period | Primary duration [year] | $N_{evt}^{primary}$ |
|---|---|---|---|---|---|---|
| California | 961,240 | ANSS | 3.5 | 1985/01/01 – 2023/11/02 | 38.80 | 8,265 |
| New Zealand | 694,460 | GeoNet | 4.0 | 1990/01/01 – 2023/12/12 | 33.94 | 9,762 |
| CSES | 781,240 | CENC | 3.0 | 2000/01/01 – 2023/08/23 | 23.63 | 9,163 |

$M_{co}^{best}$: the best or smallest magnitude of completeness above which earthquake data are considered complete; CSES: the China Seismic Experimental Site. The areas extend beyond the political borders. For more details about the catalog source, please refer to the Data and Resources section.

**Table S2.** ETAS model parameters calibrated form observed catalogs at Mcobest and utilized for simulating seismicity in the study regions.

| | Scenario | $N_{bkg}$ | $\Phi$ | $\log_{10}K$ | $\alpha$ | $\log_{10}d$ | $1+\rho$ | $\gamma$ | $\log_{10}c$ | $1+\omega$ | $\log_{10}\tau$ | $\log_{10}D$ | $Q$ |
|---|---|---|---|---|---|---|---|---|---|---|---|---|---|
| **California** | 1 | 2,517.73 | −6.73 | −0.41 | 1.03 | −0.39 | 0.69 | 1.34 | −2.70 | 1.04 | 3.84 | 1.20 | 0.64 |
| | 2 | 2,894.03 | −6.67 | −0.50 | 1.11 | −0.39 | 0.70 | 1.35 | −2.69 | 1.04 | 3.20 | 1.20 | 0.68 |
| | 3 | 2,423.85 | −6.75 | −0.39 | 1.11 | −0.40 | 0.70 | 1.36 | −2.71 | 1.03 | 3.88 | 1.20 | 0.62 |
| | 4 | 2,772.60 | −6.86 | −0.39 | 1.05 | −0.42 | 0.64 | 1.40 | −2.69 | 1.03 | 3.90 | 1.20 | 0.54 |
| **New Zealand** | 1 | 3324.34 | −6.41 | −0.51 | 1.42 | 1.18 | 0.97 | 0.69 | −2.21 | 1.14 | 3.58 | 2.02 | 1.03 |
| | 2 | 3430.56 | −6.40 | −0.56 | 1.45 | 1.18 | 0.97 | 0.69 | −2.21 | 1.14 | 3.28 | 2.00 | 1.00 |
| | 3 | 3258.05 | −6.42 | −0.48 | 1.44 | 1.18 | 0.98 | 0.68 | −2.22 | 1.14 | 3.63 | 2.04 | 1.03 |
| | 4 | 3530.52 | −6.62 | −0.47 | 1.40 | 1.23 | 0.92 | 0.63 | −2.24 | 1.13 | 3.61 | 1.97 | 0.83 |
| **CSES** | 1 | 3012.10 | −6.35 | −0.54 | 0.95 | 1.20 | 2.48 | 0.54 | −3.01 | 0.88 | 2.83 | 1.70 | 0.79 |
| | 2 | 3049.79 | −6.34 | −0.57 | 1.00 | 1.20 | 2.49 | 0.54 | −3.03 | 0.87 | 2.79 | 1.69 | 0.79 |
| | 3 | 2588.77 | −6.42 | −0.38 | 0.91 | 1.19 | 2.31 | 0.53 | −3.00 | 0.89 | 3.00 | 1.90 | 0.87 |
| | 4 | 2966.04 | −6.56 | −0.37 | 0.90 | 1.19 | 2.03 | 0.49 | −3.05 | 0.86 | 3.03 | 1.81 | 0.59 |

**Note:** Parameters $c$ and $\tau$ are given in days; $\Phi$ is the background rate per km$^2$ per year on a logarithmic scale.